# Focusing light through dynamical samples using fast closed-loop wavefront optimization

B. Blochet[1,2,*], L. Bourdieu[2], S. Gigan[1]

[1] Laboratoire Kastler Brossel, UPMC-Sorbonne Universités, ENS-PSL Research University, CNRS, Collège de France ; 24 rue Lhomond, F-75005 Paris, France
[2] IBENS, Département de Biologie, Ecole Normale Supérieure, CNRS, Inserm, PSL Research University, F-75005 Paris, France
*Corresponding author: baptiste.blochet@lkb.ens.fr

**We describe a fast closed-loop optimization wavefront shaping system able to focus light through dynamic scattering media. A MEMS-based spatial light modulator (SLM), a fast photodetector and FPGA electronics are combined to implement a closed-loop optimization of a wavefront with a single mode optimization rate of 4.1 kHz. The system performances are demonstrated by focusing light through colloidal solutions of TiO2 particles in glycerol with tunable temporal stability.**

In the last years, efforts have been made to increase the penetration depth of microscopy in tissues. The penetration depth is mostly limited by scattering, which results from optical index inhomogeneities [1, 2]. The intensity of ballistic photons (i.e. non-scattered photons) decreases exponentially during the light propagation in scattering samples following the Beer-Lambert law. Different approaches have been used to tackle this issue, in particular in non-linear microscopies. Imaging with higher wavelength [3] is less sensitive to scattering. Using regenerative amplifiers increases the amount of photons per pulse at constant average power, such that the limit of fluorescence detection is reached deeper in the tissue [4, 5]. Adaptive optics allows correcting system and sample aberrations to increase the imaging depth and resolution [6]. Despite these technical developments, imaging in tissues is still limited to superficial layers. Indeed, the bottleneck is the limited number of ballistic photons, which decay exponentially as a function of depth. In contrast, the number of scattered photons only decreases linearly with depth. Therefore, an imaging system relying on scattered photons could theoretically image much deeper in tissues.

In the past decade, wavefront shaping techniques were developed to focus light through or inside a static scattering media [7, 8], essentially by controlling interferences of the scattered light. However, biological tissues are not static, as shown by the temporal decorrelation of the speckle pattern over a large range of timescales, between 1 ms and 1 s [9]. To be able to focus light at depth in tissues, wavefront shaping needs thus to be achieved faster that these decorrelation times. Digital optical phase conjugation (DOPC) [10, 11] can be completed in a few milliseconds. However DOPC can so far only focus light inside a medium with an acoustic feedback and the size of the focus is limited [12]. Holographic techniques [13] can achieve fast focusing in few dozens of milliseconds with an optical resolution, but the imaging remains limited at intermediary depths, where light is not completely multiply scattered. Iterative wavefront optimization [14] has no fundamental limitations in reaching the optical resolution or in depth but is intrinsically slow, and technical developments are requested to accelerate the process [15, 16, 17]. In these systems, one iteration takes long between 0.12 to 10 ms.

Iterative wavefront shaping uses feedback algorithms, which consist on the selection of a SLM spatial mode (one or many pixels), the application of phase shifts between 0 and $2\pi$ for this mode and the determination of the phase that maximizes a merit factor of the intensity pattern (e.g. the intensity at a given position to focus light). By repeating this procedure for different modes, a focus is created. We can distinguish two broad approaches; one consists in displaying a series of wavefront, measuring the result, and at the end of the process, calculates and displays the optimal phase mask to obtain a focus, which can be described as an "open loop" approach. Another broad class of algorithms continuously optimizes the wavefront, depending on the result of the measurement, analogous to a "closed-loop" approach. Open loop algorithms [7, 18] may suffer from low signal to noise when the initial signal is low, but do not require fast electronic feedback. While requiring slightly more steps, a closed-loop algorithm has several advantages: it generates a focus very rapidly, and the signal to noise ratio increases iteration after iteration, resulting progressively in a better estimation of the optimum phase. In addition, the performance obtained with closed-loop algorithms do not depend on the number of SLM pixels used, thus no prior knowledge of the dynamics properties of the medium is needed to adjust the number of modes [18]. Other approaches based on genetic algorithms are also successful in creating a focus but are typically slower and more computationally demanding.

Here we report a fast closed-loop wavefront optimization system based on high-speed MEMS-based spatial light modulator (SLM), FPGA electronics and a fast mono-detector. For a $2\pi$ modulation, the SLM has a refresh rate up to 10 kHz at 532 nm, which is two orders of magnitude faster than liquid crystal phase modulators. This speed is sufficient to follow the typical decorrelation timescales of biological samples. We show here that we are able to focus light through dynamic scattering media with similar temporal [9] and scattering properties [19] as those of biological tissues.

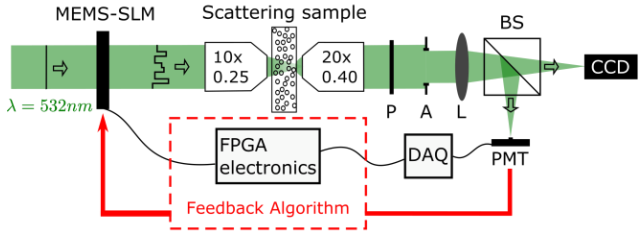

Fig. 1. Experimental setup. P: polarizer; A: aperture; L: lens (focal length = 150mm); BS: beamsplitter; MEMS-SLM; MEMS-based spatial light modulator. The wavefront of a collimated laser beam (532 nm) is modulated by a phase-only spatial light modulator. The phase mask is imaged on the back aperture of a microscope objective and focused onto a scattering sample. A second microscope objective images the output speckle using a beamsplitter on a CCD camera and on a PMT. The PMT collects the intensity of one speckle grain through an optical fiber. An iris controls the aperture size to match the speckle grain size with the diameter of the fiber. A polarizer selects one polarization state of the output speckle. The PMT signal is acquired by a DAQ board and sent to the FPGA board. During the optimization algorithm, the FPGA board computes the optimal phase for a given Hadamard mode, adds it to the current phase mask of the SLM and applies the new mask on the SLM.

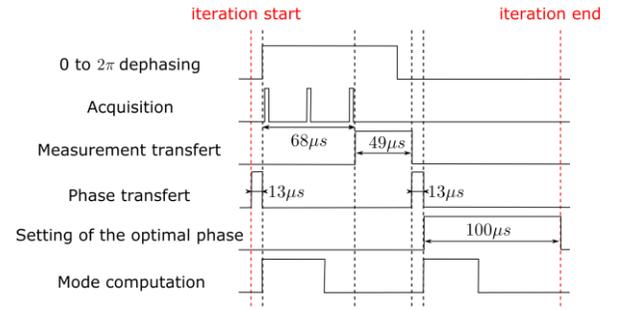

Fig. 2 Workflow of one iteration during the FPGA-based optimization wavefront shaping. At the beginning of a new iteration, a phase mask is sent to the SLM. A continuous phase difference between 0 and $2\pi$ is first imposed at all the pixels of one Hadamard mode (consisting of half of the total pixels of the SLM). On the flight, 3 measurements are acquired corresponding to phases of 0, $2\pi/3$ and $4\pi/3$. The 3 measurements are transferred to the FPGA board, which computes the optimum phase value for this Hadamard mode and adds it to the previous phase mask of the SLM. The new optimum mask is then displayed on the SLM in 100 μs. In parallel, the Hadamard mode for the next iteration is computed.

Fig. 1 describes the experimental wavefront shaping setup. The wavefront of a CW laser $\lambda$=532 nm (Coherent Sapphire) is modulated by a phase-only MEMS spatial light modulator (Kilo-DM segmented, Boston Micromachines). The SLM is conjugated to the back focal plane of a microscope objective (10x, 0.25), which illuminates a scattering sample (static or dynamic). A second microscope objective (20x, 0.4) images the output speckle onto a CCD camera (Allied Vision Technologies Manta_G-046B) and on a mono-detector (PMT, Hamamatsu H10721-20). The PMT is conjugated with the same plane and collects the intensity of one speckle grain through a multimode fiber (Thorlabs M64L02), as a feedback for our optimization process. We matched the speckle grain size to the diameter of the fiber by adjusting the aperture size of the system with an iris. One polarization state of the output speckle is selected with a polarizer.

A closed-loop algorithm is used and works as follows. At each iteration, a Hadamard mode consisting of half of the SLM pixels is selected. A continuous phase shift from 0 to $2\pi$ is imposed at these pixels, in addition to the previous phase mask applied on the SLM. The half other pixels stay fixed, as a reference. The PMT signal is acquired when the phase on the SLM is 0, $2\pi/3$ and $4\pi/3$. A 3-phase shifting interferometry algorithm then computes the phase that maximizes the PMT intensity. A new optimum phase mask is finally generated by adding to the previous mask the Hadamard mode with this optimum phase. This mask is applied on the SLM before starting the next iteration. When the full Hadamard basis has been optimized, the algorithm restarts with the first mode of this basis. The process is finally stopped when a stable gain in intensity (see below) has been reached. The CCD camera is only used for passive monitoring of the process.

A FPGA board (NI PXIe-7962R) controls all the optimization. The PMT intensity measurements are collected by a DAQ board (NI PXIe-6361), before being transferred to the FPGA board. The optimization algorithm is implemented on the FPGA board, which directly writes onto the SLM the desired phase mask. Another independent DAQ acquires the PMT intensity to monitor the process without slowing it down. A sketch of the workflow is shown on Fig. 2 for the optimization of one mode. At the beginning of a mode iteration, a phase mask is transferred onto the SLM in 13 μs. The wavefront is shifted in 100 μs from 0 to $2\pi$ continuously and the FPGA sends 3 triggers, such that the DAQ board acquires 3 values of the PMT intensity, corresponding to the phases 0, $2\pi/3$ and $4\pi/3$. After the end of the third measurement (i.e. after 68 μs), the data are transferred from the DAQ board to the FPGA board in 49 μs. After a negligible computation time for the 3-phase shift interferometry algorithm, the optimum phase mask is sent to the SLM in 13 μs, and requires an additional 100 μs to be effectively applied. This last step guarantees that our SLM will be in his linear region (phase shift between 0 and $2\pi$) for the next iteration. As the phases applied to correct the wavefront are randomly distributed between 0 and $2\pi$, a phase shift closed to $2\pi$ is generally applied at least on a few pixels. In parallel, the Hadamard vector for the next iteration is computed. To conclude, one mode optimization takes 243 μs (i.e. a "one-mode optimization rate" of 4.1 kHz). It corresponds, equivalently, to display 3 phase masks and one optimal mask at a refresh rate of 16 kHz.

To validate the optimization algorithm and its workflow, we first used this system to focus a laser beam through a static scattering sample (ground glass, Thorlabs DG10-120 A, see Fig. 3). The intensity enhancement η, defined as the ratio of the intensity at the focus after optimization to the mean speckle intensity, characterizes the efficiency of the optimization: η = $I_{focus}/I_{speckle}$. In a static sample, the maximum enhancement is proportional to the number of pixels of the SLM [8]. After a few hundred milliseconds of optimization, a plateau is reached, which means that all the modes are fully optimized. When the plateau is reached, the intensity at the focus varies significantly due to the optimization process, which generates almost 100% modulation. While a single optimization already allows reaching an enhancement of 120, it requires 3 to 5 iterations across the Hadamard basis to reach its maximum, of the order of 210, which is a typical value for an optimization performed with 1020 pixels. On Fig. 3b, the main steps of the optimization can be identified: the learning step (blue line), the application of the optimum phase (green line) and the main transfer times (red line).

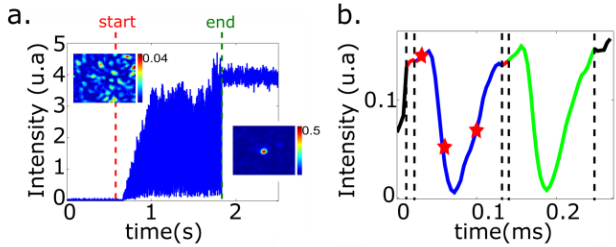

Fig. 3 Optimization through a static sample (ground glass). (a) Optimization of 5000 modes. Red dotted line: start of the optimization. Green arrow: end of optimization. The two insets show the CCD images before and after optimization. For this sample, an enhancement of 210 is obtained. (b) Zoom on the optimization of one mode. The red lines correspond to the transfer from or to the FPGA. The blue line corresponds to the learning step and the red stars to the 3 intensity measurements. A mode optimization ends by the display of the optimum phase (green line), which in general consists in a phase difference close to $2\pi$. The optimization of one mode takes 243 µs.

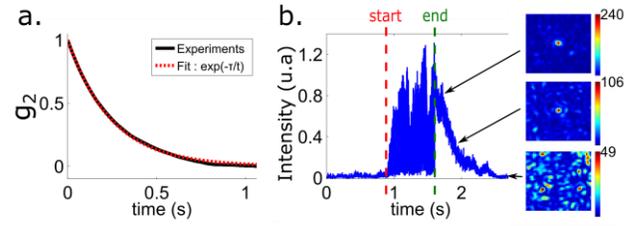

Fig. 4 Optimization through a dynamic sample. (a) Autocorrelation of the speckle measured with the CCD camera. By fitting the autocorrelation function with an exponential, the characteristic decorrelation time of the speckle is measured at the output of the sample. For this solution $\tau$ = 250 ms. (b) Focusing through a dynamic scattering media. Before the beginning of the optimization (red dotted line), the speckle is fluctuating in time due to the Brownian motion of $TiO_2$ particles in glycerol. When the optimization starts, the intensity rapidly increases until it reaches a plateau. An intensity enhancement of 80 is obtained. The optimization ends at the green line. Evolution of the focus on the CCD is shown after the end of the optimization.

Focusing through a dynamic scattering media requires now to generate the focus before the medium has changed. The stability of the medium is characterized by the characteristic decorrelation time $\tau$ of the speckle, i.e. the mean time during which the output speckle can be considered as static. To focus through scattering media, it is therefore necessary to optimize enough modes to get a significant intensity enhancement during this characteristic time. For fast decorrelating samples, the effective number of modes that contributes to the focus is proportional to the decorrelation time $\tau$ divided by the time spent to optimize one mode $T_{mode}$. The enhancement, which is proportional to the number of mode, should be proportional to the ratio $\tau/T_{mode}$. Therefore we expect that the enhancement increases linearly with $\tau$ for small decorrelation times, and eventually saturates for more stable samples to a value similar to the one obtained with static media [8, 17, 18]. Finally, the times required to obtain a stable focus should also be of the order of $\tau$, to reach the steady state between optimization of the focus and decorrelation of the speckle.

To test our system on well-controlled dynamic samples, we designed colloidal scattering solutions of $TiO_2$ particles in glycerol with tunable temporal stability. Tuning the temperature changes the glycerol viscosity, and thus the dynamic properties of our samples due to Brownian motion. This Brownian motion results in a close to exponential decay of the speckle intensity auto-correlation $g_2$ [20], which time constant $\tau$ is the characteristic decorrelation time of the speckle.

We used a solution of $TiO_2$ particles (Sigma Aldrich 224227) in glycerol with a mass concentration of 20g/l and a thickness of 500 µm. This solution has a scattering strength ($l_s$ = 70 µm and $l^*$ = 200 µm) similar to the one of biological tissues [19]. Tuning the temperature between 10°C and 24°C changes the decorrelation time of this sample from 350 ms to 30 ms, also within the characteristic range of decorrelation times of biological samples [9].

In Fig. 4a, an exponential fit of the autocorrelation function indicates a decorrelation time of 250 ms for this solution at 15°C. Fig. 4b shows an example of the optimization through this sample. After a few characteristic decorrelation times, the optimization reaches an equilibrium, when decorrelation and optimization equalize. The resulting intensity enhancement (~80) is reduced compared to the one obtained with a static sample. Even considering only the points where the optimal phase of each mode is applied, some temporal fluctuations of the optimized focus are still observed due to phase measurement errors, and decorrelation. When the optimization ends, due to the dynamic of the medium, the focus intensity decays with a timescale comparable to the decorrelation time of the medium.

In Fig. 5.a, we measured the intensity of the optimized focus for 3 samples with different characteristic decorrelation times, tuned between 30 ms and 340 ms. Results were averaged over 25 realizations. The optimizations reached a plateau for these three different samples. This plateau is obtained within a characteristic time of the order of their decorrelation time. Accordingly, the intensity enhancements decreased, as expected, with the stability of the sample. In Fig. 5b, a nearly linear relationship between the enhancement and the decorrelation time is evidenced.

For large decorrelation times (340 ms), an enhancement of 110 is reached, close to the one obtained in the static regime after a single optimization. This is consistent with the fact that the full Hadamard basis is optimized at least once during the decorrelation time. The saturation of the enhancement at large decorrelation times could not be however evidenced due to the impossibility in practice to prepare samples at lower temperatures. On the other hand, for the sample with the fastest dynamics (30 ms), the optimization reaches a typical enhancement of 10, which is still large and might be enough to obtain a significant signal for imaging. Let us note that images could be then obtained by scanning the focus thanks to the memory effect [21]. As the focus has a lifetime of the order of the decorrelation time, an image has to be obtained within this timescale [9].

The system described in this paper is to our knowledge, the fastest optimization wavefront shaping with a closed-loop algorithm described in the literature. Compared to previous closed-loop implementations, one mode is optimized 37 times faster [17]. A faster setup, albeit open loop, has been described in the literature [16] (0.125 ms/mode). As a proof of principles, we have shown that our setup allows refocusing scattered light within fast decorrelating samples, with timescales of decorrelation as small as 30 ms. This opens new possibility to refocus light efficiently in biological tissues despite their dynamical properties characterized by decorrelation times that vary from the millisecond to the second timescale.

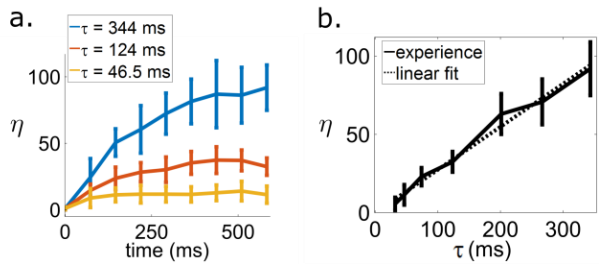

Fig. 5 Optimization versus decorrelation time. For the same scattering sample, by tuning the temperature, the decorrelation time changes from 46.5 ms to 344 ms. (a) Evolution in time of the intensity at the focus (average over 25 optimizations and standard deviation). The faster the speckle decorrelation, the faster the focus optimization reaches its plateau value, and the lower the enhancement. (b) Maximum intensity enhancement versus decorrelation time. The enhancement is proportional to τ.

Some key features of our system are critical to obtain these performances. The frequency cut off of our PMT electronics is set to 33 kHz to suppress as much noise as possible without attenuating the $2\pi$ modulation. In our experiments, the SNR is sometimes very low, of the order of unity. Still, the system is able to find a strong optimized focus. In this context, the use of a closed-loop algorithm is advantageous, since the mask application at each iteration increases the SNR at each iteration. Along the same idea, dephasing half of all the pixels at each iteration using the Hadamard basis increases the SNR. However, if a strong error affect the estimation of the optimal phase for a mode and is applied, the focus will be strongly reduced. To avoid this effect, smarter algorithms could be implemented, where for instance the number of pixels per mode decreases towards the end of the optimization.

A solution to accelerate the optimization would be to overdrive the SLM. Sending a larger phase target to the micro-mirror and changing the target phase at the appropriate time will stop the micro-mirrors at the desired value with an improved temporal resolution (Meadowlark, SLM ODP). However, more complex algorithms might incur a penalty in term of computation time and might reduce the speed.

To summarize, we have successfully used a MEMS-based SLM coupled to FPGA electronics to implement a fast optimization wavefront shaping. We were able to focus through a scattering media with a rate of 4.1 kHz per mode. The system fulfills the requirements in term of speed and enhancement to focus through biological tissues. Experiments were performed in transmission through a scattering sample and a linear signal is used as a feedback for the algorithm. However, the method described here is fully compatible with an epidetection and with any feedback signal. In particular, to achieve the same optimization inside a sample, a non-linear feedback could be used [14].

**Funding**. European Research Council (ERC) (278025 and 724473); This work has received also support under the program « Investissements d'Avenir » launched by the French Government and implemented by ANR with the references ANR–10–LABX–54 MEMOLIFE and ANR–10–IDEX–0001–02 PSL*. B. B. was funded by a PhD fellowship from Université Pierre et Marie Curie under the program «Interface pour le Vivant».

**Acknowledgment**. The authors thank Markus Ludwig for the help on the temperature controller, David Martina and Toufik Elatmani for the help on the electronics and Cathie Ventalon for fruitful discussions. S. G. is a member of the Institut Universaire de France.

**REFERENCES**
1. F. Helmchen and W. Denk, Nat. Methods 2, 932 (2005).
2. V. Ntziachristos, Nat. Methods 7(8), 603 (2010).
3. D. G. Ouzounov, T. Wang, M. Wang, D. D. Feng, N. G. Horton, J. C. Cruz-Hernández, Y.-T. Cheng, J. Reimer, A. S. Tolias, N. Nishimura, and C. Xu, Nat. Methods 14(4), 388 (2017).
4. P. Theer, M.T. Hasan, and W. Denk, Opt. Lett. 28, 1022 (2003).
5. E. Beaurepaire, M. Oheim, and J. Mertz, Opt. Commun. 188, 25 (2001)
6. N. Ji, D. E. Milkie, and E. Betzig, Nat. Methods 7(2), 141 (2010).
7. S. M. Popoff, G. Lerosey, R. Carminati, M. Fink, A. C. Boccara, and S. Gigan, Phys. Rev. Lett. 104(10), 100601 (2010).
8. I. M. Vellekoop and A. P. Mosk, Opt. Lett. 32(16), 2309 (2007).
9. Y. Liu, P. Lai, C. Ma, X. Xu, A. A. Grabar, and L. V. Wang, Nat. Commun. 6, 5904 (2015).
10. Y. Liu, C. Ma, Y. Shen, J. Shi and L.V. Wang. Optica, *4*(2), 280. (2017)
11. D. Wang, E. H. Zhou, J. Brake, H. Ruan, M. Jang, and C. Yang, Optica 2(8), 728 (2015).
12. R. Horstmeyer, H. Ruan, and C. Yang, Nat. Photonics 9, 563 (2015).
13. I. N. Papadopoulos, J.-S Jouhanneau, J. F. A. Poulet, and B. Judkewitz, Nat. Photonics 11, 116 (2017).
14. O. Katz, E. Small, Y. Guan, and Y. Silberberg, Optica 1(3), 170 (2014).
15. M. Cui, Opt. Lett. 36(6), 870 (2011).
16. A. S. Hemphill, J. W. Tay, and L. V. Wang, J. Biomed. Opt. 21, 121502 (2016).
17. C. Stockbridge, Y. Lu, J. Moore, S. Hoffman, R. Paxman, K. Toussaint, and T. Bifano, Opt. Express 20(14), 15086 (2012).
18. I. M. Vellekoop and A. P. Mosk, Opt. Commun. 281(11),
19. S. L. Jacques, Phys. Med. Biol. 58(11), R37 (2013).
20. P. Apollo, Y. Wong, and P. Wiltzius, Rev. Sci. Instrum. 64(9), 2547 (1993).
21. I. M. Vellekoop and C. M. Aegerter, Opt. Lett. 35(8), 1245 (2010).